\documentclass[pdflatex,sn-mathphys-ay]{sn-jnl}

\usepackage[T1]{fontenc}
\usepackage[utf8]{inputenc}
\usepackage{graphicx}%
\usepackage{multirow}%
\usepackage{amsmath,amssymb,amsfonts}%
\usepackage{mathtools,dsfont}
\usepackage{amsthm}%
\usepackage{mathrsfs}%
\usepackage[title]{appendix}%
\usepackage[dvipsnames]{xcolor}%
\usepackage{textcomp}%
\usepackage{manyfoot}%
\usepackage{booktabs}%
\usepackage{algorithm}%
\usepackage{algorithmicx}%
\usepackage{algpseudocode}%
\usepackage{listings}%
\usepackage{tikz}

\theoremstyle{thmstyleone}%
\theoremstyle{thmstyletwo}%

\theoremstyle{thmstylethree}%
\theoremstyle{remark}

\newcommand{\1}{\mathbf{1}}

\newcommand{\N}{\mathbb N}

\newcommand{\calH}{\mathcal H} 
\newcommand{\calV}{\mathcal V} 
\newcommand{\calE}{\mathcal E} 
\newcommand{\calP}{\mathcal P} 

\newcommand{\smax}{s_{\textrm{max}}}

\DeclareGraphicsExtensions{.pdf, .jpg, .jpeg, .png, .gif,.eps}
\graphicspath{{figs/}{.}}

\definecolor{uniformblue}{RGB}{0, 102, 204} %
\definecolor{lightgreen}{RGB}{102, 255, 102} 
\definecolor{lightorange}{RGB}{255, 180, 102}

\begin{document}

\title[Statistical modeling of hypergraphs]{
A statistical perspective on higher-order interactions modeling
}

\author*[1,2,3]{\fnm{Catherine} \sur{Matias}}\email{catherine.matias@math.cnrs.fr}

\affil*[1]{\orgname{Sorbonne Université}, \orgaddress{\postcode{F-75005}, \city{Paris}, \country{France}}}

\affil[2]{\orgname{Université Paris Cité}}

\affil[3]{\orgdiv{Laboratoire de Probabilités, Statistique et Modélisation}, \orgname{CNRS}}


\abstract{
Modeling higher-order interactions (HOI) has emerged as a crucial challenge in complex systems analysis, as many phenomena cannot be fully captured by pairwise relationships alone. Hypergraphs, which generalize graphs by allowing interactions among more than two entities, provide a powerful framework for representing such intricate dependencies. Adopting a statistical and probabilistic perspective on hypergraph modeling, we propose a guided tour through this emerging research area.  

We begin by illustrating the ubiquity of HOI in real-world systems, where interactions often involve groups of entities rather than isolated pairs. We then introduce the foundational concepts and notations of hypergraphs, discussing their descriptive statistics, graph-based representations, and the challenges associated with their complexity. We further explore a variety of  statistical models for hypergraphs and address the critical task of node clustering. We conclude by outlining some open challenges in the field. 
}

\keywords{hypergraph model, node clustering, }



\maketitle


\section{Introduction}

The growing interest in modeling higher-order interactions (HOI) arises from the acknowledgement that many phenomena are fundamentally more complex than what pairwise relationships alone can capture. 
While networks and their mathematical representation as graphs capture interactions between pairs of entities, HOI are inherently of a different nature, as they may involve the interaction of more than two elements. 
Taking into account HOI offers a richer and more expressive way to model complex interactions across diverse fields, ranging from social network analysis \citep[early acknowledged in][]{simmel_1902_1,simmel_1902_2} 
or co-authorship relations~\citep{roy:ravi:15} to ecological systems \citep{muyi:20}, neurosciences \citep{chela:21} or chemistry \citep{restrepo26}, among others.

Recent reviews on HOI include \cite{batt:etal:20,bick:etal:21,torr:etal:21}
and mostly focus on the complex systems point of view from physics. 
We choose  to focus on the \emph{statistical modeling and probabilistic point of view} of HOI \citep[also adopted in][]{lee_survey_2025} and  will mostly focus on hypergraphs.

\paragraph*{What this review is not about.}
HOI analysis comes after a data-collection step, in which HOI could either be directly observed, or \emph{inferred} from preliminary data \citep[e.g.][]{lizotte_hypergraph_2023}. The construction or the inference of these HOI is not discussed here.
Simplicial complexes~\citep{Bianconi_2021} are often presented as an alternative to hypergraphs for modeling HOI. These come with node positions in a topological space, a feature that could reveal quite useful. However 
\emph{valid} structures impose a nestedness property, where every subset of interacting entities is assumed to be interacting. 
While this assumption may be appropriate for e.g. for proximity interactions (see next section), in most applications this appears too restrictive (the co-authorship exemple being the most proeminent situation where this assumption is not appropriate). Moreover, even in cases where this assumption might not be a strong constraint, one might question  the appropriateness of introducing this supplementary information into the modeling (for instance because it might introduce additional noise). An interesting exploration about the level of \emph{simpliciality} (i.e. the inclusion structure) of HOI may be found in \cite{landry2024}. 
To keep our contribution relatively concise, neither dynamics on HOI nor temporal aspects of HOI will be covered here. 
Finally, Bayesian hypergraphs are probabilistic models where dependencies of a set of random variables is described by HOI, generalizing Bayesian networks \citep{javidian_hypergraph_2020}. This topic is thus not concerned with \emph{observed HOI} on which we focus here.

\section{Examples of systems showing HOI}
Borrowing from the approach of Holme in his review of temporal networks \citep{Holme_review}, we start by a quick guided tour on dataset types and more generally on systems where HOI naturally occur. Interestingly, an important part of the systems cited by Holme appear to be HOI in their raw format, subsequently reduced to pairwise interactions. Rather than providing an exhaustive list of datasets or publications where HOI appear, we stress the potential ubiquity of these type of data.  
Notice also that any bipartite network naturally produces a HOI, or said differently many HOI have been considered up to now as bipartite networks. We discuss in Section~\ref{sec:HOI2graph} the differences between these two approaches.

\paragraph*{Social Sciences and Ethology.}
Social interactions are of primary interest and motivated a vast majority of the modeling developments in network science. While dyadic interactions are the simplest, 
early acknowledgement of the role and importance of larger interactions appeared in the Sociology literature \citep{simmel_1902_1,simmel_1902_2}. 
Humans and animals (separately) are the classical entity sets considered in social interactions. Now, most of these interactions are either sampled as raw HOI or may be naturally constructed from raw data in the same way as pairwise interactions did. This is the case in particular for radio-frequency identification data where individual positions are recorded and interactions occur between entities lying within a ball of a given radius; observations from the field where  humans / animals gathering are recorded\footnote{see for e.g \url{https://sociopatterns.org/}}.
Communications may include the classical email exchanges (with multiple receivers) or conference calls (between humans) as well as non-verbal group interactions \citep{Social_groups}. 
Scientific collaboration (e.g. co-authorship) are probably  the most proeminent example of HOI~ \citep[e.g.][]{battiston_higher-order_2025,roy:ravi:15}, and is also the perfect toy example to explain the difference between HOI and the clique of all pairwise interactions. Variants of these data include software development as published in web platforms~\citep{Rust}. 
A broader approach to collaboration involves systems where individuals form a HOI when they serve together on the same company board \citep{aksoy}.
In the same way, the now classical ``Les Misérables'' dataset describing how the characters from Victor Hugo's novel interact in the different  scenes of the book has been studied from the HOI point of view \citep{aksoy}. The same can be done for actors playing in movies (e.g. with data extracted from the Internet Movie DataBase).

\paragraph*{Natural Sciences.}
Neuroscience and connectomics is an important source of HOI data inside human brain, with recent approaches relying on functional magnetic resonance imaging data \citep{santoro_higher-order_2024} or electro- and  magneto-encephalograms signals \citep{bilbao_higher-order_2026}.
HOI also appear when considering genetic disorders, with genes mutations implicated in a specific disease  \citep{aksoy} or metabolic pathways where  interacting entities are metabolites \citep{HyperEmbed}.
More generally, HOI are used in chemistry to describe components involved in a chemical reaction \citep{flam:etal15}.
Ecology has  seen a surge in attention towards HOI, around the idea that most pairwise interactions are in fact mediated by additional actors \citep{Malyon23,Mayfield}. 


\section{Concepts, notation and representations of HOI}
We start this section by providing the basic definitions around the concept of hypergraphs, that will be our standard representation of HOI. We then continue with graph representations of HOI, emphasizing their limitations. 
\subsection{Hypergraphs}

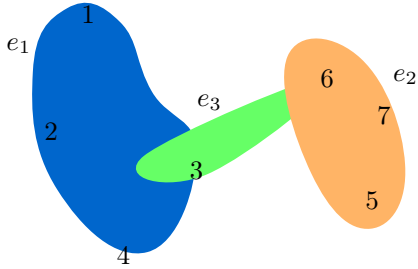
\begin{figure}
\centering
\begin{tikzpicture}[scale=1,rotate=-60]
    \tikzstyle{vertex}=[circle,fill=black,minimum size=5pt,inner sep=0pt]
    \tikzstyle{hyperedge}=[draw=blue!50, fill=blue!10, ultra thick, rounded corners=20pt, inner sep=10pt]

    \node[vertex] (v1) at (0,0) {};
    \node[vertex] (v2) at (1,-1) {};
    \node[vertex] (v3) at (2,0) {};
    \node[vertex] (v4) at (2.5,-1) {};
    \node[vertex] (v5) at (3.5,1.8) {};
    \node[vertex] (v6) at (2,2) {};
    \node[vertex] (v7) at (3,2.5) {};
    
    \fill[uniformblue, thick, fill opacity=0.2, draw=uniformblue]    
	plot [smooth cycle, tension=0.8] coordinates {(0,0.3)  (-0.5,-0.2)  (0,-1) (1.5,-1.5)  (2.8,-0.9) (2,0.3) (1,0.2)};

    \fill[lightgreen, fill opacity=0.2, draw=lightgreen, thick]
        plot [smooth cycle, tension=0.8] coordinates {(2,2.3) (2.3,0.3) (1.7,-0.3)};

    \fill[lightorange, fill opacity=0.2, draw=lightorange, thick]
        plot [smooth cycle, tension=0.8] coordinates {(1.5,2) (3,1.5) (4,1.8) (3.5,2.6) (2,2.8)};

    \node at (-0.3,0) {1};
    \node at (0.8,-1.2) {2};
    \node at (2.2,0.2) {3};
    \node at (2.7,-1.2) {4};
    \node at (3.7,2) {5};
    \node at (2,2.3) {6};
    \node at (2.8,2.7) {7};
    
    \node at (-0.4,-1) {$e_1$};
    \node at (2.5,3.2) {$e_2$};
 \node at (1.5,0.8) {$e_3$};
\end{tikzpicture}

\caption{A hypergraph with 7 nodes and 3 hyperedges: $e_1=\{1,2,3,4\}$, $e_2=\{5,6,7\}$ and $e_3=\{3,6\}$.}
\label{fig:toy_hypergraph}
\end{figure}


A hypergraph (Fig.~\ref{fig:toy_hypergraph}), denoted $\calH=(\calV, \calE)$, comprises a set of (undistinguishable) nodes $\calV=\{1,\dots, n\}$ and a set of hyperedges $\calE \subset \calP(\calV)$, where $\calP(\calV)$ is the set of all subsets of $\calV$. In other words,  each hyperegde $e\in \calE$ is a subset of nodes in $\calV$ 
and represents an interaction between those entities. 
The \emph{order} of $\calH$ is its number of nodes $|\calV|=n$;  while its \emph{size} is its number of hyperedges $|\calE|=M$. 

The simplest hypergraphs are binary (hyperedges record the presence/absence of interactions between subsets of nodes) but may  be generalized to  \emph{multiple} (or weighted) interactions. Then the hypergraph $\calH=(\calV, \calE, w)$ comes with a weight function $w:\calP(\calV) \to \N\cup \{0\}$ such that $\forall e \notin \calE$, we have  $w(e)=0$, and $w(e)\in \N^\star$ otherwise. 
The weight counts how many times a hyperedge appears in the hypergraph. Multiple hypergraphs can be viewed as hypergraphs where the set of hyperedges $\calE$ is allowed to be a multiset (i.e. some hyperedges may appear several times). 
A binary hypergraph is a particular case of a weighted hypergraph with weight function  being the indicator function $w(e)=\1\{  e\in \calE\}$ (i.e., each hyperedge has multiplicity 1).

The \emph{incidence matrix}  $H$ of the hypergraph has  dimension $|\calV|\times |\calE|$  and entries $H(v,e)=\1\{ v \in e\}$. 
A hypergraph is said to be \emph{$s$-uniform} if it only contains hyperegdes of cardinality $s$ (also called the hyperedge \emph{size}), in which case, it can be represented through a \emph{tensor} matrix $A\subset \calV^s$ with dimension $s$ and entry indexed by $(i_1,\dots, i_s)$ given by $\1\{\{i_1,\dots, i_s\} \in \calE\}$. A graph is a particular case of a 2-uniform hypergraph with $A$ being the classical adjacency matrix.
Sometimes hyperedges are allowed to be multisets, in which case a same node may be involved several times (i.e. with some \emph{multiplicity}) in a same hyperedge. We call these \emph{multiset hypergraphs}. For example a self-loop $\{v,v\}\in \calE$ is a multiset hyperedge with size 2.

\paragraph*{Descriptive statistics on hypergraphs.}
Some of the concepts introduced to describe graphs find a direct generalization in hypergraphs, while other, because of the increased complexity of hypergraphs versus graphs, induce more variability in their definitions. This is the case for the density. A basic definition would simply count the number of hyperedges divided by the maximum number of such, thus introducing 
\[
d(\calH) = \frac{|\calE|}{\sum_{s=2}^{\smax} \binom n s},
\]
where $\smax$ is the largest hyperedge size observed. Note that such a definition implicitly assumes that hyperedges of size larger than $\smax$ are impossible.  
A more refined definition would consider that each hypergraph $\calH=(\calV,\calE)$ is a collection of $s$-uniform hypergraphs $\calH_s=(\calV,\calE_s)$ over a common set of nodes $\calV$, thus introducing  the sequence  $(d_s=d_s(\calH_s))_{s \ge 2}$ of the  frequencies of hyperedges with size $s$, namely 
\[ d_s(\calH_s) = \frac {|\calE_s|} {\binom n s}.
\]
Other variants could be designed, relying on  averages over hyperedge sizes and measuring slightly different characteristics of the data.  

In contrast to this flexibility and variety, the degree of a node is simply the number of hyperedges it belongs to: $\deg_{\calH}(v) =\sum_{e \in \calE} \1\{ v \in e\}$;  while the size of a hyperedge $e$ is the number of nodes it contains: $|e| =\sum_{v\in \calV} \1\{ v \in e\}$.  
Node degrees (resp. hyperedge sizes) correspond to the row (resp. column) sums of the incidence matrix $H$. A weighted version with  entries $H(v,e)=w(e)\1\{v \in e\}$ gives rise to \emph{weighted node degrees} obtained as row sums,  and \emph{weighted hyperedges sizes} obtained by column sums. 

Centrality measures rely on the notion of \emph{paths} and describe the propensity of a node (or an interaction) to
 be 
such that any information flow passing between 2 random nodes in the system will (frequently) pass through that node (or interaction).
A $k$-path is a (finite) sequence of hyperedges  where 2 successive elements share at least $k$ common nodes, with $k=1$ being the weakest notion (in force in the context of graphs). Note that introducing a width overlap $k$ is crucial to capture the higher-order aspect of those structures. This further gives rise to $k$-distances between two nodes, defined by the smallest length of any $k$-path between them~\citep{aksoy}. Centrality measures can then be defined from these distances. 

In the graph statistics literature, an important role  is played by the concepts of \emph{transitivity} or \emph{clustering  measures}. These are  inherently based on the notion of pairwise interactions, as they quantify the propensity that ``a friend of your friend is your friend''. Such concepts do not have a natural generalization in the hypergraph world \citep[though some tentative definition exist, see for e.g.][]{transitivity}. 
Nonetheless these quantities are also linked with the frequency of ``triangles'' (i.e. cycles with length 3) and moving to the more general concept of \emph{motif} frequencies, one may naturally generalize these to the hypergraph context, with the only limitation of the increasing complexity in the variety of motifs \citep{juul_motifs,lotito_motifs}.

\textbf{Large-scale hypergraphs characteristics.} 
Whereas in the early 2000s, a large body of literature explored the characteristics of real graphs on a large scale, leading to the formulation of general laws such as the degrees scale-free distribution or the small-world property, such large scale exploration  has received little attention up to now. This could be either due to the computational complexity of these data or a potentially larger diversity of the structures that would prevent from the emergence of general rules. On a moderate scale, we mention that \cite{do2020,Lee21} have explored the characteristics  of thirteen real-world hypergraphs from various domains, with a focus on the overlaps of hyperedges for the latter reference. 

\textbf{Complexity.} 
While the number of possible edges in a graph grows quadratically with the number of nodes, the number of possible hyperedges in a hypergraph grows exponentially with that number. Indeed, 
a (simple) hypergraph with $n$ nodes may contain at most $\sum_{s=2}^n \binom{n}{s} = 
2^n -n-1$ hyperedges. This raises non trivial challenges from the statistical inference point of view and one possible approach to addressing this issue is mentioned in Section~\ref{sec:LSM} when discussing the work by~\cite{Fritz26}.

\subsection{Graph representations} 
\label{sec:HOI2graph}
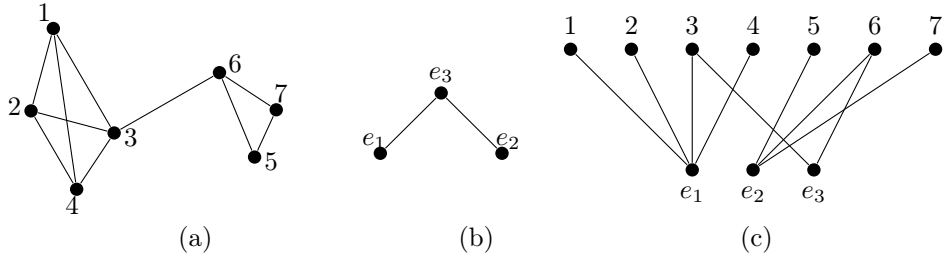
\begin{figure}
\centering
\hspace{-1cm}
\begin{minipage}{0.3\textwidth}
\begin{tikzpicture}[scale=0.8, rotate=-60]
    \tikzstyle{vertex}=[circle,fill=black,minimum size=5pt,inner sep=0pt]

    \node[vertex] (v1) at (0,0) {};
    \node[vertex] (v2) at (1,-1) {};
    \node[vertex] (v3) at (2,0) {};
    \node[vertex] (v4) at (2.5,-1) {};
    \node[vertex] (v5) at (3.5,1.8) {};
    \node[vertex] (v6) at (2,2) {};
    \node[vertex] (v7) at (3,2.5) {};
    
    \draw (v1) -- (v2);
    \draw (v1) -- (v3);
    \draw (v1) -- (v4);
    \draw (v2) -- (v3);
    \draw (v2) -- (v4);
    \draw (v3) -- (v4);
    \draw (v3) -- (v6);
    \draw (v5) -- (v6);
    \draw (v5) -- (v7);
    \draw (v6) -- (v7);

    \node at (-0.3,0) {1};
    \node at (0.8,-1.2) {2};
    \node at (2.2,0.2) {3};
    \node at (2.7,-1.2) {4};
    \node at (3.7,2) {5};
    \node at (2,2.3) {6};
    \node at (2.8,2.7) {7};
    
\end{tikzpicture}
\end{minipage}
\hspace{0.5cm}
\begin{minipage}{0.3\textwidth}
\begin{tikzpicture}[scale=0.8]
    \tikzstyle{vertex}=[circle,fill=black,minimum size=5pt,inner sep=0pt]
%
    \node[vertex] (e1) at (0,0) {};
    \node[vertex] (e3) at (1,1) {};
    \node[vertex] (e2) at (2,0) {};

    \draw (e1) -- (e3);
    \draw (e3) -- (e2);

    \node at (-0.1,0.2) {$e_1$};
    \node at (1,1.3) {$e_3$};
    \node at (2.1,0.2) {$e_2$};
    
\end{tikzpicture}
\end{minipage}
\hspace{-1.5cm}
\begin{minipage}{0.4\textwidth}
\begin{tikzpicture}[scale=0.8]
    \tikzstyle{vertex}=[circle,fill=black,minimum size=5pt,inner sep=0pt]
%
    \node[vertex] (v1) at (-1,2) {};
    \node[vertex] (v2) at (0,2) {};
    \node[vertex] (v3) at (1,2) {};
    \node[vertex] (v4) at (2,2) {};
    \node[vertex] (v5) at (3,2) {};
    \node[vertex] (v6) at (4,2) {};
    \node[vertex] (v7) at (5,2) {};
    \node[vertex] (e1) at (1,0) {};
    \node[vertex] (e2) at (2,0) {};
    \node[vertex] (e3) at (3,0) {};

    \draw (e1) -- (v1);  \draw (e1) -- (v2); \draw (e1) -- (v3); \draw (e1) -- (v4);
    \draw (e2) -- (v5); \draw (e2) -- (v6);  \draw (e2) -- (v7);
    \draw (e3) -- (v3);  \draw (e3) -- (v6);   

    \node at (-1,2.4) {1};
    \node at (0,2.4) {2};
    \node at (1,2.4) {3};
    \node at (2,2.4) {4};
    \node at (3,2.4) {5};
    \node at (4,2.4) {6};
    \node at (5,2.4) {7};

    \node at (1,-0.4) {$e_1$};
    \node at (2,-0.4) {$e_2$};
    \node at (3,-0.4) {$e_3$};

\end{tikzpicture}
\end{minipage}
\\ 
\hspace{-1cm} (a) \hspace{3cm} (b) \hspace{3cm} (c)
\caption{Graph representations of the hypergraph from Fig.~\ref{fig:toy_hypergraph}. (a) Clique graph; (b) Line graph; (c) Bipartite graph.}
\label{fig:graphs}
\end{figure}

Due to their complexity, it is tempting to reduce hypergraphs to simpler objects such as graphs (see Fig.~\ref{fig:graphs} for an illustration), that are easier to handle. However this is at the cost of either loosing information or relaxing some constraints, as we now explain. 

\paragraph*{Clique graph.}
The clique graph of a hypergraph (also called 2-section, clique expansion or clique reduction) has the same set of nodes, and edges between nodes that share a hyperedge. Each hyperedge $e\in \calE$ in the hypergraph is in fact \emph{reduced} into a complete clique in the graph. A weighted version can also be used, transferring partial information about the hyperedges sizes to the projected graph.  
In any case, this naive representation looses a lot of information and it is impossible to reconstruct hyperedges from the clique graph. 

\paragraph*{Line graph.}
The line graph of a hypergraph has vertices corresponding to the hyperedges of that hypergraph,  and edges between overlapping hyperedges (\emph{i.e.} that share at least one node). Again, this representation looses information (about how many and which nodes are shared) and (unique) recovery of a hypergraph from its line graph is not possible. The line graph is mostly used to summarize adjacency relations between hyperedges (two hyperedges being adjacent when they share a node).

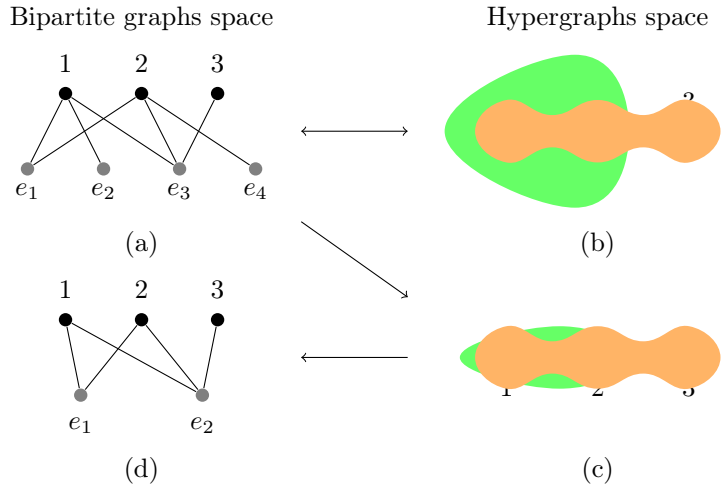
\begin{figure}
\centering
\begin{tikzpicture}[
    squig/.style={decorate, decoration={snake, amplitude=1mm, segment length=4mm}}
]
\tikzstyle{vertex}=[circle,fill=black,minimum size=5pt,inner sep=0pt]
\tikzstyle{hype}=[circle,fill=gray,minimum size=5pt,inner sep=0pt]

\node at (0,3.5) {Bipartite graphs space};

\node[vertex] (v1) at (-1,2.5) {};
\node[vertex] (v2) at (0,2.5) {};
\node[vertex] (v3) at (1,2.5) {};

\node[hype] (s1) at (-1.5,1.5) {};
\node[hype] (s2) at (-0.5,1.5) {};
\node[hype] (s3) at (0.5,1.5) {};
\node[hype] (s4) at (1.5,1.5) {};

\draw (v1)--(s1);
\draw (v1)--(s2);
\draw (v1)--(s3);

\draw (v2)--(s1);
\draw (v2)--(s3);
\draw (v2)--(s4);

\draw (v3)--(s3);

    \node at (-1,2.9) {1};
    \node at (0,2.9) {2};
    \node at (1,2.9) {3};
    \node at (-1.5,1.2) {$e_1$};
    \node at (-0.5,1.2) {$e_2$};
    \node at (0.5,1.2) {$e_3$};
    \node at (1.5,1.2) {$e_4$};

    \node at (0,0.5) {(a)};


\node[vertex] (a2) at (-1,-0.5) {};
\node[vertex] (b2) at (0,-0.5) {};
\node[vertex] (c2) at (1,-0.5) {};
\node[hype] (t1) at (-0.8,-1.5) {};
\node[hype] (t2) at (0.8,-1.5) {};

\draw (a2)--(t1);
\draw (b2)--(t1);
\draw (b2)--(t2);
\draw (c2)--(t2);
\draw (a2)--(t2);

    \node at (-1,-0.1) {1};
    \node at (0,-0.1) {2};
    \node at (1,-0.1) {3};
    \node at (-0.8,-1.9) {$e_1$};
    \node at (0.8,-1.9) {$e_2$};

\node at (0,-2.5) {(d)};

\draw[<->] (2.1,2) -- (3.5,2);
\draw[<-] (2.1,-1) -- (3.5,-1);
\draw[->] (2.1,0.8) -- (3.5,-0.2);


\node at (6,3.5) {Hypergraphs space};

\node[vertex] (ha) at (4.8,2) {};
\node[vertex] (hb) at (6,2) {};
\node[vertex] (hc) at (7.2,2) {};

    \node at (4.8,2.4) {1};
    \node at (6,2.4) {2};
    \node at (7.2,2.4) {3};

 \fill[red, fill opacity=0.35, draw=red, thick]
    (ha) circle (0.3); 
            
 \fill[lightgreen, fill opacity=0.2, draw=lightgreen, thick]
    plot [smooth cycle, tension=0.8] coordinates {(4,2) (5.7,3)  (6.4,2) (5.7,1) };
    
    \fill[lightgreen, fill opacity=0.2, draw=lightgreen, thick]
        plot [smooth cycle, tension=0.8] coordinates {(4.2,2) (5.5,2.4)  (6.4,2) (5.5,1.6) };

\fill[lightorange,fill opacity=0.2, draw=lightorange, thick]
 plot [smooth cycle, tension=0.8] coordinates {(4.4,2) (4.8,2.4) (5.4,2.2) (6,2.4) (6.6,2.2) (7.2,2.4) (7.6,2) (7.2,1.6) 
 (6.6,1.8) (6,1.6) (5.4,1.8) (4.8,1.6) };

\node at (6,0.5) {(b)};


\node[vertex] (ha2) at (4.8,-1) {};
\node[vertex] (hb2) at (6,-1) {};
\node[vertex] (hc2) at (7.2,-1) {};
    \node at (4.8,-1.4) {1};
    \node at (6,-1.4) {2};
    \node at (7.2,-1.4) {3};

    \fill[lightgreen, fill opacity=0.2, draw=lightgreen, thick]
        plot [smooth cycle, tension=0.8] coordinates {(4.2,-1) (5.5,-0.6)  (6.4,-1) (5.5,-1.4) };

\fill[lightorange,fill opacity=0.2, draw=lightorange, thick]
 plot [smooth cycle, tension=0.8] coordinates {(4.4,-1) (4.8,-0.6) (5.4,-0.8) (6,-0.6) (6.6,-0.8) (7.2,-0.6) (7.6,-1) (7.2,-1.4) (6.6,-1.2) (6,-1.4) (5.4,-1.2) (4.8,-1.4) };

\node at (6,-2.5) {(c)};

\end{tikzpicture}

\caption{(a) A bipartite graph $\mathcal{G}$; (b) Projection of $\mathcal{G}$ into the space of multisets hypergraphs  with self-loops, choosing the top nodes of $\mathcal G$ as the new set of nodes. Hyperedges are $\{1,2\}, \{1\}, \{1,2,3\}$ and $\{1,2\}$. The applications from (a) to (b) are invertible bijections, one being the inverse of the other; 
(c) Projection of $\mathcal{G}$ on the simple hypergraphs subspace: the multiplicity of hyperedge $\{1,2\}$ and the self-loop $\{1\}$ have been removed. (d) Embedding of the simple hypergraph from (c) in the bipartite graphs space. Note that (a) and (d) are not the same bipartite graph.}
\label{fig:bipartite}
\end{figure}

\paragraph*{Bipartite graph.}
A more elaborate graph representation of a hypergraph consists in considering its bipartite 
representation (or star-expansion graph), in which the hypergraphs nodes form a first nodes part, while the set of hyperedges forms the second nodes part. An edge in the bipartite graph is drawn from an original node to an original hyperedge (now a second part's node) whenever it belongs to it in the hypergraph. Under some conditions, this is a lossless process. More precisely, given a (simple) bipartite graph and the choice of one part as the original set of nodes, one can reconstruct a unique (multiset) hypergraph over this set of nodes, which may eventually contain multiple hyperedges and self-loops 
(see Fig.~\ref{fig:bipartite}). In other words, bipartite graphs may be embedded into a general space of hypergraphs and simple hypergraphs may be projected into bipartite graphs.

\section{Statistical models of hypergraphs}

\subsection{Randomness is in the hyperedge: limitations with models on bipartite graphs}
From the previous section, it  seems natural to use bipartite graph models in order to derive hypergraph models. 
However, this might be done only at some additional cost, as we now explain. 

Random graphs models always consider the set of nodes $\calV$ as deterministic and focus on the randomness in the links, aka the edges in the graph. In particular, the number of such links is most often random, excepted for   the  Erdős-Rényi  variant $\mathcal{G}(n,M)$, where the number $M$ of edges is fixed and their locations (among the $\binom n 2$ pairs of nodes) are random. This variant is  asymptotically equivalent  to the $\mathcal{G}(n,p)$ one (where all edges appear independently with probability $p$) in that if  $M=M_n$ and $p\in (0,1)$ satisfy $|M_n-\binom n 2 p| =O(n\sqrt{p(1-p)})$, then  if the probability of an event $E$ tends to some $c\in [0,1]$ under the distribution $\mathcal{G}(n,M_n)$, it also converges to the same value under the distribution  $\mathcal{G}(n,p)$ \citep[see][]{Luczak87}.  

Now, a statistical model over the bipartite representation of a hypergraph will also always consider a fixed set of bipartite nodes, resulting in a fixed number of hyperedges in the hypergraph. The only randomness we can get lies in which nodes are involved in each of the  $M$ interactions, corresponding to the randomness in the formation of the links in the bipartite graph.  Contrarily to the $\mathcal{G}(n,M)$ case, fixing the number of hyperedges in a model does not in general lead to an  asymptotically equivalent reformulation of another model  with  random number of hyperedges.  As a consequence, hypergraph models derived from bipartite graphs are not the most general.  For instance, a hypergraph stochastic blockmodel (SBM, see Section~\ref{sec:SBM} below) is more general than the corresponding bipartite SBM formulation, as the latter imposes a group structure on the hyperedges \citep[see Section A3 in the Supp. Mat. of][]{HyperSBM}.

\subsection{Uniformly random, configuration, and preferential attachment models} 
Generalizing the Erdős-Rényi random graph model yields uniformly random hypergraphs. This approach involves uniformly sampling from the set of all $s$-uniform hypergraphs  
defined over a set of $n$ nodes. However, much like the Erdős-Rényi model for graphs, this hypergraph model is overly simplistic and homogeneous for meaningful statistical analysis of real-world datasets.

The exponential random graph approach led to the proposal of a $\beta$-model for hypergraphs \citep{stasi}, where hyperedges occur independently, and the sufficient statistic of the model is the degree sequence (or nodes degrees specific to the hyperedges sizes). This model was further theoretically studied in \cite{nandy}.

Configuration models for random graphs consist in uniformly sampling from the set of all possible graphs over $n$ nodes, while adhering to a prescribed degree sequence. For hypergraphs, these were first introduced by \cite{ghos:zlat:cald:etal:09}, focusing on tripartite and 3-uniform hypergraphs. Later, \cite{chod:20} extended this framework to the non-uniform case. In these works, both node degrees and hyperedge sizes remain fixed—a consequence of relying on bipartite representations of hypergraphs.
The configuration model is particularly valuable for sampling graphs (resp. hypergraphs) that match the degree sequence (resp. and hyperedge sizes) of an observed dataset, typically via shuffling algorithms. As such, it is frequently employed as a null model in statistical analyses. However, exact sampling (as opposed to approximate sampling) from this model presents significant challenges, especially for hypergraphs~\cite[see Section 4 in][for more details]{chod:20}.

Preferential attachment (PA) models have been proposed in~\cite{Wang_evolving}, where both the idea of hyperedge growth and hyperedge preferential attachment were introduced. Those ideas were later refined in \cite{Guo} and more recently in~\cite{jung2026}. 
\cite{Barthelemy22} proposes a very general formulation for the probability that a vertex belongs to an edge, which thus boils down  to relying on the bipartite graph representation. His approach comprises Erdős-Rényi-like, configuration, (a sort of) PA, and random geometric models. The latter models are designed to be generative and are not readily amenable to statistical inference.

\subsection{Latent space and block models}
\label{sec:LSM}
Latent space models (LSM) for hypergraphs raise the issue of constructing a proximity indicator or measure for a subset of more than 2 latent positions. 
\cite{Turnbull_etal} propose a random geometric hypergraph model, where hyperedges form between nodes as soon as latent-position balls of some radius intersect. To avoid imposing a simplicial complex structure, the radii differ by node subset size, increasing with it and thus preventing automatic inclusion of smaller subsets. This deterministic framework is then augmented with a random step, though this introduces identifiability challenges, which are mitigated via prior distributions during inference.
\cite{lyu2023} proposed a tensor-based LSM, however limited to 3-uniform hypergraphs. 
Proximity measures for subsets of nodes may also rely on averages of their (relative) latent positions (e.g. arithmetic, geometric, Hölder, $\ldots$). This is the avenue pursued in \cite{Fritz26}, and combined with a latent hyperbolic space, taking advantage of a more expressive geometry towards hierarchical and embedded structures. That work also contains a most promising tool for the statistical analysis of hypergraphs: a sample-to-population estimation procedure, that consists in replacing the model likelihood by an approximation where non-occurring interactions are only sampled while the occurring ones (the hyperedges) are all included.  

Block models will be discussed in Section~\ref{sec:SBM}, as their discrete latent space is directly linked to node clustering. We mention here the work by \cite{ng:murp:22} that proposes a mixture model on the hyperedges and is thus not linked to node clustering. Finally, \cite{bala:21} proposes a nonparametric hypergraphon model, but limited to the uniform case.

\section{Node clustering on hypergraphs}
\label{sec:cluster}

\paragraph*{What cluster types are we looking for?} 
The simplest type of cluster is a community, characterized in the context of graphs by groups of nodes that are strongly connected internally but weakly connected externally. The first question that arises is: What is a community in a hypergraph? One could consider that hyperedges constitute (overlapping) communities,  in which case clusters are directly observed. A more refined definition would state that nodes  that often share the same hyperedges form a community. A key challenge here is wether the size of those hyperedges should be taken into account or not ? For instance, is there a community structure in the toy hypergraph from Fig.~\ref{fig:toy_hypergraph}? There, node groups  $\{1,2,3,4\}$ and $\{5,6,7\}$ have as many internal as external hyperedges (one, respectively) but the sizes of the internal hyperedges are larger.
In fact, a wide variety of definitions are possible, giving rise to equally diverse proposals in the literature. 

\paragraph*{Can we hope to detect them?}
The node clustering issue is intimately linked to the existence of information-theoretic limits that prevent from recovering or detecting those clusters.
That question has been initially approached  in the context of uniform hypergraphs, thereby limiting the scope of the results. Indeed, though hypergraphs may be seen as a collection of $s$-uniform hypergraphs for varying values of $s$, it is not necessary that all layers be informative to recover the underlying latent structure. 
Non uniform results in sparse hypergraphs include \cite{zhen2023} which contains  convergence bounds for both the model parameters and the communities, and  \cite{dumitriu2025} that provides a weak consistency result on the communities when model parameters are known. 
More recently, \cite{ruggeri2024_MP} established a first detection result valid in a non-uniform hypergraph, however restricted to a particular setting where the probability of a hyperedge is expressed as the sum of pairwise probabilities. This \emph{dyadic} restriction makes the model more similar  to the graph setting.

\paragraph*{Model based approaches - SBM}
\label{sec:SBM}
Beyond communities, the blockmodel approaches simply define clusters as groups of nodes with same (conditional) interaction probabilities. 
Many proposals have emerged in the literature these past few year, together with 
degree-corrected variants~\citep{ghos:dukk:14,Chodrow_21,yuan2022,HyperSBM}.

\paragraph*{Other approaches.}
Other approaches to node clustering in hypergraphs include modularity-based methods. Modularity definitions heavily rely on the definition of a community and various directions have been followed in that area. 
The reader will find a comparison of these methods in~\cite{modularity}. 
An alternative is provided by spectral clustering. Most existing methods heavily rely on the (weighted) clique graph representation \citep{ghos:dukk:17:aos}, at the cost of loosing information, while others are either restricted to simplicial structures or uniform hypergraphs~\cite{delgenio25}.
Finally, some approaches based on random walks have been suggested~\cite{swan:zhan:21}.



\section{Conclusions and next challenges}

Scalability is certainly one of the most challenging issue in hypergraphs modeling. It comes in two ways: being able to handle potentially large hyperedges sizes in one hand, and more generally large systems (in the number of individuals and interactions) in the other hand. Approximate inference is certainly a promising avenue in that direction, as initiated for instance in \cite{Fritz26}. Efficient softwares for statistical analysis need to be developed, in line with the existing libraries such as HyperNetX~\citep{hypernetx}.
Impossibility results or phase transition thresholds for community detection in non-uniform hypergraphs seem difficult to obtain and are certainly one of the next challenges in this area. As already stressed, the uniform hypergraphs results won't help in that direction as not all layers need to be informative. Moreover, the only available threshold~\citep{ruggeri2024_MP} heavily relies on a \emph{dyadic type} modeling assumption.
Synthetic benchmark data for community detection in hypergraphs are urgently needed. These may not rely on hypergraph SBM, so that the model-based methods are not favored in the comparison with the others. Some non convincing proposals have been made \citep[see the discussion in][]{modularity}, and again, this raises the delicate question of community definition in the hypergraph context.

\backmatter

\bmhead{Acknowledgements}
I deeply thank the organizers of the workshop ``New Trends in Statistical Network Analysis'' during which the idea of this special issue arose, namely Carsten Jentsch, Göran Kauermann and Alexander Kreiss, especially for fostering fruitful and engaging exchanges among all participants.  

\section*{Statements and Declarations}

\begin{itemize}
\item Funding 
Not applicable 
\item Competing interests
Not applicable
\item Ethics approval and consent to participate
Not applicable
\item Consent for publication
Not applicable
\item Data availability
Not applicable 
\item Materials availability
Not applicable
\item Code availability 
Not applicable
\item Author contribution
Not applicable
\end{itemize}


\bibliography{Hypergraphs2026}

\end{document}